\documentclass[showpacs,preprintnumbers,amsmath,amssymb]{revtex4}

\usepackage{amsmath}
\usepackage{graphicx}
\usepackage{epsf}
\usepackage{psfrag}
\usepackage{epsfig}
\usepackage{graphics}

\begin{document}

\newcommand{\be}{\begin{equation}}
\newcommand{\ee}{\end{equation}}
\newcommand{\bq}{\begin{eqnarray}}
\newcommand{\eq}{\end{eqnarray}}
\newcommand{\dt}{\frac{d^3k}{(2 \pi)^3}}
\newcommand{\dtp}{\frac{d^3p}{(2 \pi)^3}}

\title{{\bf On the equivalence between Implicit Regularization and
Constrained Differential Renormalization}}

\date{\today}

\author{Carlos R. Pontes$^{(a)}$} \email []{crpontes@fisica.ufmg.br}
\author{A. P. Ba\^eta Scarpelli$^{(b)}$} \email[]{scarp1@des.cefetmg.br}
\author{Marcos Sampaio$^{(c)}$}\email[]{marcos.sampaio@th.u-psud.fr}
\author{J. L. Acebal$^{(b)}$} \email[]{acebal@dppg.cefetmg.br}
\author{M. C. Nemes$^{(a)}$}\email[]{carolina@fisica.ufmg.br}

\affiliation{(a) Federal University of Minas Gerais -
Physics Department - ICEx \\ P.O. BOX 702, 30.161-970, Belo Horizonte
MG - Brazil}
\affiliation{(b) Centro Federal de Educa\c{c}\~ao Tecnol\'ogica - MG \\
Avenida Amazonas, 7675 - 30510-000 - Nova Gameleira - Belo Horizonte
-MG - Brazil}
\affiliation{(c) Universit\'e Paris XI - Laboratoire de Physique
Th\'eorique \\
B\^atiment 210 \,\,  F-91405 \,  Orsay \, Cedex. }

\begin{abstract}
\noindent
Constrained Differential Renormalization (CDR) and the constrained
version of Implicit Regularization (IR) are
two regularization independent techniques that do not rely on dimensional continuation of the space-time. These two
methods which have rather distinct basis have been successfully applied to several
calculations which  show that they can be trusted as practical, symmetry invariant frameworks
(gauge and supersymmetry included) in perturbative computations even beyond one-loop order.

In this paper, we show the equivalence between these two methods at
one-loop order. We show
that the configuration space  rules of CDR can be mapped into the momentum space procedures of Implicit
Regularization, the
major principle behind this equivalence being the extension of the
properties of
regular distributions to the regularized ones.

\end{abstract}


\maketitle

\section{Introduction}

\indent
The problem of a simple regularization technique from the
calculational point of view  that respects gauge invariance and
supersymmetry is still of great relevance, especially beyond one-loop level. The most simple and
pragmatical regularization scheme known is Dimensional
Regularization. It is gauge invariant, since the manifest gauge
symmetry is not spoiled by the dimensional modification of the
amplitude. This is so because in this dimensional modification, all
the properties of regular integrals are retained, like the vanishing
of the surface terms and the preservation of the vector algebra (see
the section 3 of \cite{dimr}). Nevertheless, this is not the case when
the theory to be treated is supersymmetric. The dimensional
modification spoils the symmetry between fermions and bosons.
Dimensional Reduction \cite{dimred} appeared as a new supersymmetric
invariant version of this method. It only modifies the dimension of
the integral and preserves the fields and the other mathematical
objects in the proper dimension of the theory. Some
important steps towards a rigorous and model independent generalization of Dimensional
Reduction beyond one loop order have been given \cite {stock}, but all order statements have not been established.

Differential Renormalization (DR) \cite {df1} is a method that works in
the proper dimension of the theory
in coordinate space. It has been proved to be quite simple and powerful
in various applications
\cite{df2}, \cite{df3}, \cite{df4} \cite{df5}. The original DR
consists in the manipulation of singular distributions attributing to them
properties of the regular ones. They are expressed in terms of a simpler
singular function and then it is performed
its substitution by a renormalized one. This procedure originates an
arbitrary mass parameter for each
different expression. When symmetries are involved, relations between
these parameters are established
in order to obtain a symmetric result. The Constrained (version of) Differential
Renormalization (CDR) \cite{cdf1}
was developed
in order to automatically satisfy the symmetries without the need of
such adjustments at the end of
the calculations. For this, a set of rules was stated, which are
actually extensions of some additional properties of
regular distributions to the singular ones. A series of applications
of this technique was successfully carried out, which includes abelian
and non-abelian gauge symmetry,
supersymmetric theories and supergravity calculations
\cite{cdf2}-\cite{cdf7}.

Implicit Regularization (IR) \cite{ir1}, \cite{ir2}, \cite{ir3} is a
momentum-space regularization method defined in the physical
dimension of the underlying theory. The basic idea behind the method
is, after implicitly assuming some (unspecified) regulating function
as part of the integrand of divergent amplitudes, to extend all the
properties of regular integrals to the regularized ones. An algebraic identity
is used to expand the integrand and separate their regularization
dependent parts from the finite one. Symmetries of the model,
renormalization or phenomenological requirements  determine
arbitrary parameters introduced by this procedure. In fact, there is
a special choice of the parameters that automatically preserves the
symmetries in all anomaly free cases we have studied
\cite{ir1}-\cite{ir16}. The possibility of these parameters being
fixed at the begining of the calculation is desirable, since it
considerably simplifies the application of the method. This results
in a Constrained (version of) Implicit Regularization (CIR).

The technique has been shown to be tailored
to treat theories with parity violating objects in integer dimensions.
This is the case of
chiral and topological field theories. The ABJ anomaly  \cite{ABJ},
and the radiative generation of a
Chern-Simons-like term,
which violates Lorentz and CPT symmetries  \cite{ir4}, \cite{ir5} are
examples where the technique was
successfully applied. Moreover the method was shown to respect
gauge invariance in both Abelian and non-Abelian theories at one-loop
order \cite{ir4}, \cite{ir5}, \cite{ir6},
\cite{ir9}. The calculation of the $\beta$-function of the
massless Wess-Zumino model (at  three loops) was also performed as a
test  of the procedure \cite{ir8}. A non-trivial test in a
supersymmetric model was performed,
in which the anomalous magnetic moment of the lepton in supergravity
was successfully calculated \cite{ir12}.
The extension of CIR to higher loop order has been implemented and has
been applied in scalar \cite{ir11} and
gauge theories \cite{ir16}. In what concerns higher order calculations,
Differential Renormalization, in its original form, has been used with
success in scalar and gauge theories.
CDR at one loop order has been used as a guide in supersymmetric
calculations at two-loop order \cite{cdf8}.

Constrained Implicit Regularization  and Constrained Differential
Renormalization (CDR) are examples of regularization methods that
work in the proper dimension of the theory. Both were shown to
respect gauge invariance in Abelian and non-Abelian theories. The
two techniques were also tested in non-trivial supersymmetric
calculations yielding positive results. Besides, although they work
in different spaces, the results are all identical. This fact
suggests the possibility of equivalence between the two frameworks.
In this paper, we show this equivalence by mapping the rules of CDR
in the ones of CIR.

The paper is divided as follows: in section II, we present the
basics of Constrained Differential Renormalization; in section III,
the basics of CIR is bypassed; the connection between the rules of
the two techniques are analyzed in section IV and, finally, the
concluding comments are presented in section V.

\section{Constrained Differential Renormalization}

\indent
We reproduce here the basics of Constrained Differential
Renormalization (CDR). Given an amplitude in position-space, it is
written as a linear combination of derivatives of basic functions.
The basic functions are products of scalar Feynman propagators with
a differential operator acting in the last one. For example, the
bubble and the triangular basic functions have, respectively, the
general form
\be
B_{m_1 m_2}[{\cal O}]=\Delta_{m_1}(x) {\cal O}_x \Delta_{m_2}(x)
\ee
and
\be
T_{m_1 m_2 m_3}[{\cal O}]=\Delta_{m_1}(x) \Delta_{m_2}(y){\cal O}_x \Delta_{m_3}(x-y),
\ee
where $\Delta_m(x)$ is the scalar Feynman propagator and ${\cal O}_x$ is a differential operator
with respect to $x$.

A crucial step in order to write the amplitude in this way is the use
of Leibniz rule for derivatives. The rules that we will list bellow permit
one
to write renormalized expressions for the basic functions such that,
when
they are substituted into the amplitude, we will have the underlying
symmetries of the theory preserved.

The rules:

\begin{enumerate}

\item Differential reduction: singular expressions are substituted by
derivatives of regular ones. For this, two steps are used:

\begin{itemize}

\item Functions with singular behavior worse than logarithmic are
reduced
to derivatives of logarithmically singular functions without
introducing any dimensionful constant.

\item For the logarithmically singular functions (at one loop) the
following
identity is used:
\be
\frac{1}{x^4}=- \frac 14 \Box \frac {\ln{x^2M^2}}{x^2}.
\ee
This relation introduces the unique mass scale of the whole process. It
plays
the role of renormalization group scale.

\end{itemize}

\item Formal integration by parts: derivatives act formally by parts
on test functions.
\be
[\partial F]^R=\partial F^R
\ee

\item Delta function renormalization rule:
\be
[F(x, x_1,\cdots, x_n)\delta(x-y)]^R=[F(x, x_1,\cdots,
x_n)]^R\delta(x-y).
\ee

\item The validity of the propagator equation:
\be
[F(x, x_1,\cdots, x_n)(\Box^x-m^2)\Delta_m(x)]^R=[F(x, x_1,\cdots,
x_n)(-\delta(x)]^R.
\ee

\end{enumerate}

With the rules above, one can find relations between the basic
functions. A table with the renormalized basic functions can always
be used to perform the calculation of any amplitude.

\section{Constrained Implicit Regularization}

\indent Implicit Regularization (IR) can be formulated by a similar
set of rules, just like CDR. The first thing to be done is writing
the momentum-space amplitude as a linear combination of basic
integrals, multiplied by polynomials of the external momentum. These
basic integrals are the Fourier transforms of the CDR basic
functions. Typical basic integrals: \be I, I_\mu, I_{\mu \nu}= \int
\frac {d^4k}{(2\pi)^4} \frac {1, k_\mu, k_\mu k_\nu}
{(k^2-m^2)[(p-k)^2-m^2]}. \ee They are, respectively, the Fourier
transforms of $B_{mm}[1]$, $B_{mm}[\partial_\mu]$ and
$B_{mm}[\partial_\mu \partial_\nu]$. As in the case of CDR, each one
of these basic integrals can be treated following a set of rules.
So, a table with their results can be used whenever a new
calculation is being performed. The rules of CIR are:

\begin{enumerate}

\item A regularization technique is applied to the integral.  It can be
maintained
implicit, but it must have some properties: it cannot modify the
integrand and the dimension of the space-time. The first property is
to preserve the finite part and the second is a requirement in order
to not violate supersymmetry. A good one would be a simple cutoff.
The problem of possible violation of symmetries by this technique
will be automatically handled by the constraining character of Implicit
Regularization;

\item The divergent part to be subtracted in a given basic integral  is
obtained
by applying recursively the identity
\bq
\frac {1}{(p-k)^2-m^2}&=&\frac{1}{(k^2-m^2)} \nonumber \\
&-&\frac{p^2-2p \cdot
k}{(k^2-m^2)
\left[(p-k)^2-m^2\right]},
\label{ident}
\eq
until the divergent part do not have the external momentum $p$ in the
denominator.
This will assure local counterterms. The remaining divergent integrals
have the general form
\be
\int_k ^\Lambda \frac{k_{\mu_1}k_{\mu_2}\cdots}{(k^2-m^2)^\alpha},
\ee
where $\int_k$ stands for $\int d^4k/(2 \pi)^4$ and the superscript
$\Lambda$
is to indicate that the integral is regularized;

\item  The divergent integrals with Lorentz indices must be expressed
in
function of surface terms. For example:
\be
\int_k ^\Lambda \frac{k_\mu k_\nu}{(k^2-m^2)^3}=
\frac 14 \left( -\int_k^\Lambda \frac{\partial}{\partial k^\nu}
\left( \frac{k_\mu}{(k^2-m^2)^2}\right) + g_{\mu \nu} \int_k^\Lambda
\frac {1}{(k^2-m^2)^2}\right).
\label{ts}
\ee

The surface terms, that vanish for integrable cases, depend here on
the regularization applied. They are symmetry violating terms.
The possibility of making shifts in the integrals needs
the surface terms to vanish. As far as loop integrals are concerned, non-null surface
terms imply that the amplitude depend on the momentum routing choice. So, the
constraint of IR is the restoring of symmetry by means of the
cancelation of these surface terms with local restoring
counterterms. In practice, we do this automatically by setting them
to zero. We will comment on the anomalous situation latter.

\item The divergent part of the integral is written in terms of the
basic divergences
\be
I_{log}(m^2)=\int_k^\Lambda \frac{1}{(k^2-m^2)^2}
\label{ilog}
\ee

and

\be
I_{quad}(m^2)=\int_k^\Lambda \frac{1}{(k^2-m^2)}.
\label{iquad}
\ee

These objects will require local counterterms in the process of
renormalization.

\end{enumerate}

Finally we can solve the finite (regularization independent) part
and define a  subtraction scheme, for instance, absorbing the basic
divergent integrals  in the renormalization constants defined by the
counterterms. This can be done in a mass independent fashion. For
this, we use a scale relation between the basic divergent integrals
which will also introduce the renormalization group scale of the
method.

In order to give an example of the use of these steps, we apply the
method to the simple logarithmically divergent one-loop amplitude
below:
\be
I=\int^\Lambda \frac {d^4k}{(2\pi)^4}
\frac{1}{(k^2-m^2)[(k+p)^2-m^2]}.
\label{B1}
\ee
By applying identity
(\ref{ident}) in the regularized amplitude above, we get
\bq
I= I_{log}(m^2)- \int_k \frac{p^2+2p\cdot k}{(k^2-m^2)^2[(k+
p)^2-m^2]},
\label{expand}
\eq
Notice that the second integral in eq. (\ref{expand}) is finite and,
because of this, we do not use the superscript $\Lambda$. It is
convenient to express the regularization dependent part, given by
(\ref{ilog}), in terms of an arbitrary mass parameter. This becomes
essential if we are treating massless theories (see ref. \cite{ir11}).
It can be done by using the regularization independent relation
\be
I_{log}(m^2)=I_{log}(\lambda^2)+b\ln{\left(
\frac{m^2}{\lambda^2}\right)},
\label{scale}
\ee
with $b=i/(4\pi)^2$. The mass parameter $\lambda^2$ is suitable to
be used as the renormalization group scale, as it can be seen in
refs. \cite{ir8}, \cite{ir11}. After solving the finite part, we are
left with
\be
I=I_{log}(\lambda^2)-bZ_0(p^2,m^2,\lambda^2),
\ee
where
\be
Z_0(p^2, m^2,\lambda^2)=\int_0^1dx\,\,\ln{\left(
\frac{p^2x(1-x)-m^2}{(-\lambda^2)}\right)}.
\ee

Finally, we would like to comment on the relation between surface terms
and anomalies.
Momentum routing invariance seems to be the
crucial property in a Feynman diagram in order to preserve symmetries.
In fact such surface terms evaluate
to zero should we employ Dimensional Regularization (DREG) to
explicitly evaluate them. This property somewhat reveals why DREG
is manisfestly gauge invariant yet it breaks supersymmetry (the
invariance of the action with
respect to supersymmetry transformations only holds in general for
specific values of
the space-time dimension) \footnote{ The idea of associating momentum
routing in the loops with
symmetry properties of the Green's functions has been exploited in a
framework named Preregularization
which did not call for momentum routing invariance but instead fixed
the routing in order to fulfill
certain Ward identities. \cite{Mackeon}}.

A particular situation, however, is the occurrence of a quantum
symmetry breaking (anomaly).
Anomalies, within perturbation theory, may present some oddities such
as preserving a certain symmetry at
the expense of adopting a special momentum routing in a Feynman diagram
e.g. in the (Adler-Bardeen-Bell-Jackiw)
AVV triangle anomaly.  In the case of chiral anomalies, IR has been
shown to preserve the democracy between
the vector and axial sectors of the Ward identities, which is a good
'acid test' for regularizations \cite{ir5}. The
arbitrary parameter represented by the surface term remains
undetermined and floats between the axial and vector sectors
of the Ward identities. That is to say, in the anomalous amplitudes,
there is
no possibility of restoring, at the same time, the axial and the
vectorial Ward
identities. The counterterm that will restore one symmetry causes the
violation of the other and, therefore,
it does not make sense to set the surface terms to zero. The answer is
to be established by physical constraints
on such amplitude. This feature has also been illustrated  in the
description of two-dimensional gravitational anomalies \cite{ir10}.

\section{Mapping Constrained Implicit Regularization in Constrained
Differential Renormalization}

\indent We show in this section that the rules of Constraining
Differential Renormalization can be mapped in the ones of
Constrained Implicit Regularization. We will sometimes reproduce
with few details calculations of reference \cite{cdf5}.

\subsection*{The rules 1 and 2 of CDR}

\indent
We begin by analyzing rules 1 and 2 of CDR. We will consider here the
simpler
two-point massless basic function, $B[1]$. The reason is that the
one-loop renormalization
of CDR will always occur when the other basic functions are written as
functions of it.
So, its renormalization is the base for finding all the other
renormalized expressions. We write
\be
B[1]=\Delta(x)1 \Delta(x)=\left(\frac{1}{4 \pi^2 x^2}\right)^2,
\ee
which after application of rule 1 gives
\be B^R[1]=-\frac 14 \left(\frac{1}{4 \pi^2}\right)^2 \Box \frac
{\ln{x^2M^2}}{x^2}.
\ee

In order to compare the two techniques, we will take this basic
function into the momentum space. The bare momentum-space expression for
$B[1]$ in Euclidian space is given by
\be
\hat B[1]=\int \frac{d^4 k}{(2 \pi)^4} \frac{1}{k^2 (p-k)^2}=I_E,
\label{IRfunction}
\ee
where $I_E=-i I$ of eq. (\ref{B1}) for the massless case.
If we intend to obtain the Fourier transform of the renormalized
expression, we will have to make use of CDR rule 2. This rule says
that we must ignore the surface term when  integration by parts
is performed. With this prescription, the derivatives act directly
on the exponential. So, we get:
\be
\hat B^R[1]=-\frac{1}{(4 \pi)^2} \ln{\left(\frac{p^2}{\bar
M^2}\right)},
\ee
with $\bar M^2=4M^2/\gamma^2$, $\gamma$ being the Euler constant.

We would like here to show that rules 1 and 2 together stands for the
subtraction of $I_{log}(\lambda^2)$ in Implicit Regularization. To
make it clear, we reproduce here the calculation of \cite{df1}, in
which the authors show that the combination of the rules 1 and 2
corresponds to the subtraction of a local cutoff dependent term. Let
us consider the exclusion of a small ball, ${\cal B}_\epsilon$, of
radius $\epsilon$. We can write
\bq
&&\hat B[1]=\int_{R^4- {\cal B}_\epsilon}d^4x
\,\,f(x)\left(\frac{1}{4 \pi^2x^2} \right)^2 =-\frac 14
\left(\frac{1}{4 \pi^2}\right)^2 \int_{R^4- {\cal B}_\epsilon}d^4x
\,\,f(x)\Box \frac {\ln{x^2M^2}}{x^2}
\nonumber \\
&&= \frac 14 \left(\frac{1}{4 \pi^2}\right)^2 \left\{
\int_{S_\epsilon} d\sigma_\mu \,\, f(x) \partial_\mu \frac
{\ln{x^2M^2}}{x^2} +\int_{R^4- {\cal B}_\epsilon}d^4x  \partial_\mu
f(x) \partial_\mu \frac {\ln{x^2M^2}}{x^2} \right\}, \label{cutoff}
\eq
with $d \sigma_\mu=\epsilon^3 \hat x_\mu \, d \hat x$ the outward
normal volume element of the 3-sphere $S_\epsilon$, which is the
boundary of ${\cal B}_\epsilon$. The second integral is well defined
and can be integrated by parts with no problem. The complete result,
taking in account that $f(x)=e^{ip \cdot x}$,
is
\bq
&&\hat B[1]=\frac{1}{(4 \pi)^2} \left(1- \ln {\epsilon^2M^2}
-\ln{\left(\frac{p^2}{\bar M^2}\right)}\right) \nonumber \\
&&=\frac{1}{(4 \pi)^2} \left(\ln {\left(\frac{\Lambda^2}{\bar
M^2}\right)}+1
-\ln{\left(\frac{p^2}{\bar M^2}\right)}\right),
\eq
where the momentum cutoff is given by $\Lambda^2=4/(\gamma^2
\epsilon^2)$.

We now remember that $\hat B[1]=I_E$ and use the Implicit
Regularization result of eq.(\ref{B1}) for $m^2=0$, so that
\be \hat
B[1]=-i\left\{ I_{log}(\lambda^2)-b
\ln{\left(\frac{p^2}{e^2\lambda^2}\right)}\right\}.
\ee
If we use a simple momentum cutoff, $\Lambda^2$, to calculate
$I_{log}(\lambda^2)$, we get
\be \hat B[1]= \frac{1}{(4 \pi)^2}
\left(\ln {\left(\frac{\Lambda^2}{e^2\lambda^2}\right)}+1
-\ln{\left(\frac{p^2}{e^2\lambda^2}\right)}\right).
\ee
It is the same result of Differential Renormalization. We just have to rescale
our mass parameter such that $e^2\lambda^2=\bar M^2$. The important
conclusion here is that the position-space surface term that is
subtracted by means of rule 2 of CDR is exactly the basic divergence
of IR, $I_{log}(\lambda^2)$.

In the eq. (2.15) of \cite{df1}, the authors find a divergent contribution in the
limit $\epsilon \rightarrow 0$, associated to the  radius $\epsilon$ of a spherical
surface around the  propagator distributional product pole. Such  a divergent counterpart
could be subtracted by adding a suitable counterterm in an arbitrary proportion to the action.
An alternative procedure would take place by considering the pole contribution as a
concentrated distribution at $x=0$. There arises a linear combination of delta functions
with arbitrary coefficients \cite{Bogoliubov} as many as higher is the pole order. In both
cases  the arbitrariness is fixed by a subtraction scheme.

\subsection*{The role of rules 3 and 4 of CDR}

\indent
Next, we dedicate ourselves to the role of rules 3 and 4 in Constrained
Differential
Renormalization in order  to understand how they are translated to Implicit
Regularization. These rules
are important in order to introduce an unique mass parameter in the
calculation that is being performed.
They turn it possible to establish relations between a basic function
with $n+1$ propagators with one with
$n$ propagators.
If this one is already renormalized, then the same mass scale is used.
Let us see how it works before
look at its version in IR.
Consider the basic function
\be
F_n[\Box-m_{n+1}^2]=\Delta_{m_1}(x_1)\cdots
\Delta_{m_n}(x_n)(\Box^{x_{1}}-m_{n+1}^2)\Delta_{m_{n+1}}(x_1+x_2+\cdots + x_n),
\ee
where $F_n[{\cal O}]\equiv F[{\cal O}](x_1,x_2,\cdots,x_n)$ and the
variables $x_1$, $\cdots$, $x_n$ are differences between
the vertex points of the loop.
With help of rule 4, we can write
\be
F_n^R[\Box-m_{n+1}^2]=-(F_{n-1}[1]\delta(x_1+x_2+\cdots+x_n))^R
\label{r4}
\ee
and, by using rule 3:
\be
F_n^R[\Box-m_{n+1}^2]=-F_{n-1}^R[1]\delta(x_1+x_2+\cdots+x_n)
\label{r3e4}
\ee
In momentum-space, a momentum $k_i$ is associated with each internal
line and therefore
with each variable $x_i$. If we consider as the loop momentum the one
associated with the last propagator, we have
\be
\hat F_{n-1}^R[{\cal O}_k](p_1,\cdots,p_{n-1})=\left(\int_k \frac{{\cal
O}_k}{(k^2+m_n^2)
[(k-p_1)^2+m_1^2]\cdots [(k-p_{n-1})^2+m_{n-1}^2]}\right)^R.
\label{n-1}
\ee
Nevertheless, Fourier transforming equation (\ref{r3e4}) yields
\bq
&&\left(\int_k
\frac{-(k^2+m_{n+1}^2)}{(k^2+m_{n+1}^2)[(k-p_1)^2+m_1]^2\cdots[(k-p_n)^2+m_n^2]}\right)^R \nonumber \\
&& = - \left(\int_k \frac{1}{(k^2+m_n^2)[(k+p_n-p_1)^2+m_1^2]\cdots
[(k+p_n-p_{n-1})^2+m_{n-1}^2]}\right)^R
\label{r3e4m}
\eq
The straightforward operation in momentum-space, which is standard in
CIR, is the simple cancelation of the factor
$(k^2+m_{n+1}^2)$, present in both numerator and denominator of the
integrand. But the sequence of procedures performed above
includes a shift $k\to k+p_n$ in the integral. It is also in accordance
with CIR, where surface terms are discarded.
Let us see how rules 3 and 4 perform this shift. We can write eq.
(\ref{r4}), in which the rule 3 has been not yet applied, in terms of
its inverse Fourier transform, as
\bq
&&F_n^R[\Box-m_{n+1}^2]= -\left(\int_{p_1,...p_{n-1}}\hat
F_{n-1}[1](p_1,\cdots,p_{n-1})
e^{ip_1\cdot x_1}\cdots e^{ip_{n-1}\cdot x_{n-1}}\delta(x_1+\cdots +
x_n)\right)^R  \nonumber \\
&&= -\left(\int_{p_1,...p_{n-1}}\int_k
\frac{1}{(k^2+m_n^2)[(k-p_1)^2+m_1^2]\cdots [(k-p_{n-1})^2+m_{n-1}^2]}
\right. \nonumber \\
&& \left. \times e^{ip_1\cdot x_1}\cdots e^{ip_{n-1}\cdot
x_{n-1}}\delta(x_1+\cdots + x_n)\right)^R.
\eq
Carrying out the Fourier transformation of the equation above and
applying rule 3, we get
\bq
&& \hat
F_n^R[-(k^2+m_{n+1}^2)]=-\int_{x_1,\cdots,x_n}\left(\int_{p_1,...p_{n-1}}\int_k
\frac{1}{(k^2+m_n^2)[(k-p_1)^2+m_1^2]\cdots [(k-p_{n-1})^2+m_{n-1}^2]}
\right. \nonumber \\
&& \left. \times e^{i(p_1-p'_1)\cdot x_1}\cdots
e^{i(p_{n-1}-p'_{n-1})\cdot x_{n-1}}e^{-ip'_n\cdot x_n}\right)^R\delta(x_1+\cdots + x_n).
\eq
Integration on $x$ variables gives us
\bq
&&\hat F_n^R[-(k^2+m_{n+1}^2)]= -\left(\int_{p_1,...p_{n-1}}\int_k
\frac{1}{(k^2+m_n^2)[(k-p_1)^2+m_1^2]\cdots [(k-p_{n-1})^2+m_{n-1}^2]}
\right. \nonumber \\
&& \left. \times \delta(p_1-p'_1+p'_n)\cdots
\delta(p_{n-1}-p'_{n-1}+p'_n)\right)^R.
\eq
This will furnish to us the same result of eq. (\ref{r3e4m}).

It is clear from the expression above that the rules 3 and 4 taken together
are equivalent,
in momentum-space, to the cancelation of
a factor $k^2+m_{n+1}^2$ (here, in euclidian space) in the numerator
with its correspondent in the denominator and the subsequent
shift $k\to k+p_n$.
A comment is in order.
There is a physical appeal for the result above. The operation we are
discussing corresponds to a point
contraction. If we consider the original outgoing external momenta
$p_1$, $\cdots$, $p_n$, we have $\sum p_i=0$.
When the point contraction is performed, the momentum $p_n$ does not
flow outward, so that the internal
momentum that circulates the loop is changed to $k+p_n$. Alternatively,
if we consider the definition
(\ref{n-1}), we have for the result of eq. (\ref{r3e4m}), $-\hat
F^R_{n-1}[1](p_1-p_n,\cdots,p_{n-1}-p_n)$. This is in accordance
with the new condition of energy-momentum conservation
$p_1+\cdots+p_{n-1}=0$.

We have seen in this calculation that the sequence of applications of
rule 3 and 4,
when observed from the momentum-space, includes a shift.
Clearly, if the integral is at least linear divergent,
this corresponds to discarding a surface term. But, as we will
show below, this is not the unique procedure of CDR that works as a
source of shifts in momentum-space. Besides,
generally (but not always) rules 3 and 4 are used with the intention of
using the renormalized version of the basic function $B[1]$,
which corresponds to a logarithmically divergent integral in
momentum-space. For this case, no surface term is missed.
As we shall see, the crucial point occurs when Lorentz indices are
involved and Leibniz rule is used.

Finally, we enforce that, in Constrained Implicit Regularization,
shifts and the cancelation of factors of
the numerator and the denominator are essential steps. With these
procedures, we can always display
the basic divergences as $I_{log}$'s and $I_{quad}$'s, which depend on
the same mass parameter.

\subsection*{Leibniz rule in position-space, ambiguities in Fourier
transforms and shifts in momentum-space}

\indent
There is an essential characteristic of Constrained
Differential Renormalization which we will show that
takes care of momentum-space surface terms: the validity of Leibniz rule. It is an essential tool
when one establishes relations between
basic functions with and without Lorentz indices. This happens in
connection with an ambiguity when
the Fourier transform of a bare basic function is performed. Let us
consider the basic function,
\be
F[\partial_\mu](x_1,x_2,\cdots,x_n)=\Delta_{m_1}(x_1)\cdots
\Delta_{m_n}(x_n)\partial_\mu^{x_1}\Delta_{m_{n+1}}(x_1+x_2+\cdots x_n),
\ee
which has the Fourier Transform
\be
\hat F[k_\mu](p_1,\cdots,p_n)=\int_{k_1,\cdots,k_{n+1}}
\frac{ik^{n+1}_\mu}{(k_1^2+m_1^2)\cdots (k_{n+1}^2+m_{n+1}^2)}
\delta(k_1+k_{n+1}-p_1)\cdots \delta(k_n+k_{n+1}-p_n).
\ee
At this point, if there is a singularity, it emerges an ambiguity:
depending on
the momentum we choose to be the loop momentum, a different momentum
routing is obtained.
In other words, the integrals will differ by a shift. First, let us
take $k_{n+1}$ to be the loop momentum. We obtain
\bq
&&\hat F[k_\mu](p_1,\cdots,p_n)= \int_k
\frac{ik_\mu}{(k^2+m_{n+1}^2)[(k-p_1)^2+m_1^2]\cdots [(k-p_n)^2+m_n^2]} \nonumber \\
&&= -\int_k \frac{ik_\mu}{(k^2+m_{n+1}^2)[(k+p_1)^2+m_1^2]\cdots
[(k+p_n)^2+m_n^2]}.
\label{tf1}
\eq
The last equality follows from the Lorentz structure of the integral.
On the other hand, if we choose $k_1$, we have
\be
\hat F[k_\mu](p_1,\cdots,p_n)= \int_k
\frac{i(p_1-k)_\mu}{(k^2+m_1^2)[(k-p_1)^2+m_{n+1}^2]\cdots [(k+p_n-p_1)^2+m_n^2]}.
\label{tf2}
\ee
It is clear that eq. (\ref{tf2}) is obtained by performing the shift $k
\to k-p_1$ in the integrand of eq. (\ref{tf1}).
There is nothing wrong with this if the integral is finite or at most
logarithmically divergent. But this is not
the case in general. If the integral is linearly divergent, for instance,
a surface term must be added to compensate
the shift. One could avoid this problem by stating, as a rule of the
technique, that the momentum
associated with the last propagator, which closes de loop, should be
the loop momentum.

Nevertheless, in some situations, if the Leibniz rule is allowed, this
ambiguity cannot be removed.
Let us consider the simple example of the
massless basic function,
\be
B[\partial_\mu]= \Delta(x)\partial_\mu \Delta(x),
\ee
which, by Leibniz rule, can be written as
\be
B[\partial_\mu]=\frac 12 \partial_\mu B[1],
\label{imu}
\ee
so that
\be
B^R[\partial_\mu]=\frac 12 \partial_\mu B^R[1].
\label{bmu}
\ee
We should call the reader's attention to the fact that eq. (\ref{imu})
was written
considering that
\be
\Delta(x)\partial_\mu \Delta(x)= \partial_\mu (\Delta(x))\Delta(x),
\ee
which, if we take into account the rule discussed above, which tells us
that the momentum
associated to the last propagator is the loop momentum, implies that
\be
-\int_k \frac{ik_\mu}{k^2(p+k)^2}=\int_k \frac{i(p-k)_\mu}{k^2(p-k)^2}.
\ee
So, we can say that the application of the Leibniz rule in
position-space, in some peculiar situations,
is equivalent to discard a surface term in momentum-space.

In the discussion that follows in this section, we will show that all
the relations between
basic functions which in CDR are obtained by using Leibniz and its 3
and 4 rules can be
obtained in momentum-space by performing shifts and by canceling common
factors of the numerator and
the denominator.
We return to equation (\ref{bmu}). If we look at this equation in
momentum-space, we have
\be
\left( \int_k \frac {k_\mu}{k^2(p-k)^2}\right)^R=\frac 12 p_\mu \left(
\int_k \frac {1}{k^2(p-k)^2}\right)^R ,
\ee
or $I_\mu=(1/2)p_\mu I$.
The relation above was obtained by the use of properties of regular
distributions
in position-space extended to singular ones. We would like to treat
directly in momentum-space
the integral $I_\mu$. We will make use of two different ways of
calculation. In the first one,
considering the extension of all the properties of regular to
regularized integrals, we perform
the shift $k \to k+p$ in the integral $I_\mu$, so that
\be
I_\mu=\int_k \frac {(k+p)_\mu}{k^2(p+k)^2}=\int_k \frac
{k_\mu}{k^2(p+k)^2}+p_\mu I= -I_\mu +p_\mu I.
\ee
In the last step, we have observed that $I_\mu$ is odd in $p$. The
equation above lead us again
to the result $I_\mu=(1/2)p_\mu I$. In the procedure above we have
shifted a linear divergent integral.
This would require the addition of a surface term. So, this step is in
accordance with the
rule of Implicit Regularization that tell us to eliminate such terms by
means of symmetry
restoring counterterms.

The second way we treat $I_\mu$ is its explicit calculation by means of
Implicit Regularization. This will
permit us to identify the forgotten surface term. We begin applying
identity (\ref{ident}) two times to
the integrand:
\be
I_\mu=\int_k^\Lambda
\frac{k_\mu}{(k^2-m^2)}\left(\frac{1}{(k^2-m^2)}-\frac{p^2-2p\cdot k}{(k^2-m^2)^2}
+\frac{(p^2-2p\cdot k)^2}{(k^2-m^2)^2[(p-k)^2-m^2]}\right),
\ee
in which we will take the limit $m^2 \to 0$ at the end of the
calculation. Eliminating the vanishing terms,
we have
\bq
&&I_\mu= 2p^\alpha \int_k^\Lambda \frac{k_\mu k_\alpha}{(k^2-m^2)^3}
+ \int_k \frac{k_\mu (p^2-2p\cdot k)^2}{(k^2-m^2)^3[(p-k)^2-m^2]}
\nonumber \\
&&=\frac {p^\alpha}{2} \left( -\int_k^\Lambda \frac{\partial}{\partial
k^\nu}
\left( \frac{k_\alpha}{(k^2-m^2)^2}\right) + g_{\mu \alpha}
\int_k^\Lambda
\frac {1}{(k^2-m^2)^2}\right)+ \tilde I_\mu,
\eq
$\tilde I_\mu$ being the finite integral. After the calculation of this
finite part,
we obtain, in the limit $m^2 \to 0$,
\bq
&&I_\mu= \frac {p_\mu}{2} \left(I_{log}(\lambda^2)-b
\ln{\left(-\frac{p^2}{\lambda^2 e^2}\right)}\right)
-\frac{p^\alpha}{2}S_{\mu \alpha}    \nonumber \\
&&=  \frac {p_\mu}{2} I -\frac{p^\alpha}{2}S_{\mu \alpha},
\eq
where
\be
S_{\mu \alpha}= \int_k^\Lambda \frac{\partial}{\partial k^\nu}
\left( \frac{k_\alpha}{(k^2-m^2)^2}\right)
\ee
is a surface term that will be set to zero and where we have made use
of the scale relation (\ref{scale}).
In the analysis above, we saw that the validity of Leibniz rule in the
calculation of $B[\partial_\mu]$
implies in the validity of a shift in a linear divergent integral. In
other words, it means that a
surface term has been subtracted. Although we are analyzing a
particular case, this relation is used
in the derivation of all basic functions with superior Lorentz indices,
as we show in some examples.

The next example we examine is the calculation of the basic function,
\be
T[\partial_\mu \partial_\nu]=\Delta(x)\Delta(y)\partial_\mu^x
\partial_\nu^x \Delta(x+y).
\ee
It can be decomposed into a
traceless and a trace part. It is added a local term to be fixed,
due to a possible ambiguity in the finite traceless part:
\be
T^R[\partial_\mu \partial_\nu]=T[\partial_\mu \partial_\nu -\frac 14
\delta_{\mu \nu} \Box] +\frac 14 \delta_{\mu \nu}T^R[\Box] +
\frac{1}{64 \pi^2}c \delta_{\mu \nu} \delta(x)\delta(y).
\ee
In the equation above the second term of the r.h.s. is renormalized by
means of rules 3 and 4 of CDR and $c$ is the arbitrary constant to
be fixed. It is fixed so that the rules of CDR are valid. Specifically
the rules of Leibniz, and 3 and 4 of CDR, states that
\be
B^R[\partial_\mu](x)\delta(y)=-\Box^y T[\partial_\mu]+2
\partial^y_\rho T^R[\partial_\mu \partial_\rho]- T^R[\Box
\partial_\mu].
\ee
The integration of this equation on $x$ yields $c=-1/2$.
We will repeat this procedure in momentum-space and show that the
constant $c$ is fixed so that it cancels the surface term that comes
from the traceless part. We have
\bq
\hat T^R[k_\mu k_\nu]&=& \hat T[k_\mu k_\nu -\frac 14 g_{\mu \nu}k^2]+
\frac 14 g_{\mu \nu}
\hat T^R[k^2]
+\frac{1}{64 \pi^2}c g_{\mu \nu}
\label{a0}
\eq
We remember that a infrared cutoff $m^2$ is used (it disappears after
the scale relation is used). By
using equation (\ref{ident}) and the identity $k^2=(k^2-m^2)+ m^2$ in
the first term, we get
\be
-\frac 14 \left\{\int^{\Lambda}_k
\frac{g_{\mu\nu}}{(k^2-m^2)^2}-
4\int^{\Lambda}_k
\frac{k_{\mu}k_{\nu}}{(k^2-m^2)^3}\right\} g_{\mu \nu} + \frac{1}{64
\pi^2}c g_{\mu \nu}+\mbox{nonambiguous terms}.
\ee
We calculate it by symmetric integration ($k_\mu k_\nu \to k^2 g_{\mu
\nu}/4$):
\bq
&&-\frac 14\left\{\int^{\Lambda}_k
\frac{g_{\mu\nu}}{(k^2-m^2)^2}-
4\int^{\Lambda}_k
\frac{k_{\mu}k_{\nu}}{(k^2-m^2)^3}\right\} + \frac{1}{64 \pi^2}c g_{\mu
\nu}\nonumber \\
&& = \frac{m^2}{4}g_{\mu \nu}\int_k \frac{1}{(k^2-m^2)^3}+\frac{1}{64
\pi^2}c g_{\mu \nu}
= -\frac{ig_{\mu \nu}}{128 \pi^2}+ \frac{1}{64 \pi^2}c g_{\mu \nu}.
\eq
So, in order to cancel this surface term, $c= i/2$, just like the
result of Constrained
Differential Renormalization (the $i$ factor is due to the fact that we
work in Minkowski space).

The two examples we have worked above are cases that involve at most
linear divergences. Besides, the number
of Lorentz indices is at most two. When the degree of divergence or the
number of Lorentz indices
increase, new surface terms appear. Let us see the case of the basic
function,
\be
B[\partial_\mu \partial_\nu]=\Delta(x)\partial_\mu \partial_\nu
\Delta(x).
\ee
Following the steps of CDR, its most general renormalized expression is
given by
\be
B[\partial_\mu \partial_\nu]= \frac 13 \left( \partial_\mu
\partial_\nu- \frac 14 \delta_{\mu \nu} \Box \right) B^R[1]
+ \frac{1}{16 \pi^2}\left[ f \partial_\mu \partial_\nu +\delta_{\mu
\nu}(g \Box + \mu^2) \right] \delta(x),
\label{a1}
\ee
where a differential equation was solved to find the first term at $x
\neq 0$ and the constants $f$, $g$ (dimensionless)
and $\mu$ (mass dimension) where introduced to take care of
ambiguities. As before, they  will be fixed
in such a way that the rules of CDR, including the Leibniz rule, are
respected.
Using these laws of manipulation, it can be shown that
\be
-\frac 12 \Box \partial_\mu B^R[1]+2 \partial_\rho B^R[\partial_\mu
\partial_\rho]=0
\label{a2}
\ee
and
\bq
&&B^R[\partial_\mu \partial_\nu](x)\delta(y)=-\Box^y T^R[\partial_\mu
\partial_\nu]
+2 \partial_\rho T^R[\partial_\mu \partial_\nu \partial_\rho]
\nonumber \\
&& +\frac 12 (\partial^x_\mu \partial^y_\nu+\partial^x_\nu
\partial^y_\mu)
\left(B^R[1](x)\delta (x+y)\right) +  B^R[\partial_\mu
\partial_\nu](x)\delta(x+y).
\label{a3}
\eq
We will also need the basic function, which in the equation bellow is
decomposed in a trace
and a traceless part, plus a local arbitrary term, left to be adjusted
according to the rules:
\bq
&&
T^R[\partial_\mu \partial_\nu \partial_\rho]= T[\partial_\mu
\partial_\nu \partial_\rho-
\frac 16(\delta_{\mu \nu}\partial_\rho+\delta_{\mu \rho}\partial_\nu
+\delta_{\rho \nu}\partial_\mu)\Box] \nonumber \\
&& +\frac {1}{12} \left(\delta_{\mu \nu}(\partial^x_\rho
+\partial^y_\rho) +
\delta_{\mu \rho}(\partial^x_\nu+ \partial^y_\nu) +\delta_{\rho
\nu}(\partial^x_\mu+\partial^y_\mu)  \right) \nonumber \\
&& \times \left( -B^R[1]\delta(x+y)+ \frac {1}{16 \pi^2}d
\delta(x)\delta(y)\right).
\label{a4}
\eq
We should notice that rules 3 and 4 and the renormalization of the basic
function $B[\partial_\mu]$ have been already applied in the trace part.
When eq. (\ref{a1}) is substituted into eq. (\ref{a2}), it is found
that $\mu=0$ and $g=-f$. If equations
(\ref{a1}), (\ref{a4}) and  (\ref{a0}) are inserted into the expression
(\ref{a3}) and integration on
$x$ is carried out, the result is: $f=\frac{1}{18}$ and $d=-\frac 13$.

Let us now see how it works in momentum-space if the principles of
Constrained Implicit Regularization
are considered. First, we show that eq. (\ref{a2}) is respected in
momentum-space, as long as shifts
in the integrand are permitted. In momentum space, we have:
\be
-\frac 12 p^2 p_\mu I + 2 \int_k^\Lambda \frac{(p\cdot
k)k_\mu}{k^2(p-k)^2}=0.
\label{a2m}
\ee
In the second integral, we can use
\be
(p \cdot k)=-\frac 12 [(p-k)^2-k^2-p^2],
\ee
so that it is given by
\be
-\left\{ \int_k^\Lambda \frac{k_\mu}{k^2} -\int_k^\Lambda
\frac{k_\mu}{(p-k)^2} -p^2 I_\mu\right \}.
\ee
The first term is obviously null. We shift the second one ($k \to k+p$)
and obtain
\be
p_\mu \int_k^\Lambda \frac{1}{k^2}  +p^2 I_\mu= p_\mu
I_{quad}(m^2=0)+\frac{p^2}{2}p_\mu I.
\ee
In the expression above, the last term cancels out the first of
eq.(\ref{a2m}). For the basic
quadratic divergence with null mass, in \cite{ir5} and \cite{ir11} it
was shown that an adequate parametrization
gives us
\be
I_{quad}(m^2)= \frac{i}{(4\pi)^2} m^2  \left[ \ln{\left(
\frac{\Lambda^2}{m^2}\right)}+ \mbox{const} \right],
\label{piquad}
\ee
so that $I_{quad}(m^2=0)=0$ and eq. (\ref{a2m}) is satisfied. We call
the
reader attention to the fact that this parametrization
furnishes the same result as the one of its correspondent in
position-space,
the one point basic function $A_m[1]=\Delta_m(x)\delta(x)$ of CDR.
The next equation to be analyzed in momentum-space is (\ref{a3}), given
by
\bq
&&-\int_k^\Lambda \frac{k_\mu k_\nu}{k^2(p-k)^2}=-p'^2 \int_k^\Lambda
\frac{k_\mu k_\nu}{k^2(p-k)^2(p'-k)^2} \\ \nonumber
&& + \int_k^\Lambda \frac{(2p' \cdot k)k_\mu k_\nu}{k^2(p-k)^2(p'-k)^2}
-\frac 12(p_\mu p'_\nu+p_\nu p'_\mu)  \int_k^\Lambda
\frac{1}{k^2(p-p'-k)^2}
-\int_k^\Lambda \frac{k_\mu k_\nu}{k^2(p-p'-k)^2}.
\label{a3m}
\eq
We begin by noting that the two first terms of the r.h.s. can be
considered together, so that we have,
in the numerator, $-p'^2+2(p'\cdot k)= k^2-(p'-k)^2$. We break it again
in two parts, and perform cancelations
with factors of the denominator. So,
\bq
&&-\int_k^\Lambda \frac{k_\mu k_\nu}{k^2(p-k)^2}=-\int_k^\Lambda
\frac{k_\mu k_\nu}{k^2(p-k)^2}      \\ \nonumber
&& \int_k^\Lambda \frac{k_\mu k_\nu}{(p'-k)^2(p-k)^2}
-\frac 12(p_\mu p'_\nu+p_\nu p'_\mu)  \int_k^\Lambda
\frac{1}{k^2(p-p'-k)^2}
-\int_k^\Lambda \frac{k_\mu k_\nu}{k^2(p-p'-k)^2}.
\eq
The three last terms of the above equation must cancel out in order to
the equation be satisfied.
If we perform the shift $k \to k-p'$ in the last two integrals and use
the result, also obtained
by means of shifts,
\be
\int_k^\Lambda \frac{k_\mu }{(p'-k)^2(p-k)^2}=
\frac{(p+p')_\mu}{2}\int_k^\Lambda \frac{1}{(p'-k)^2(p-k)^2},
\ee
the exact cancelation occurs. The reader should observe that all the
results that CDR reaches with the help
of Leibniz and its 3 and 4 rules are reached in momentum space with the
use of shifts and cancelation of
common factors of the numerator and the denominator.

We now will fix the arbitrary constants with the help of the principles
of Constrained Implicit Regularization.
First, once eq. (\ref{a2m}) was verified, it is trivial to check that
the momentum-space version of eq. (\ref{a1})
implies $f=-g$ and $\mu=0$. We turn ourselves to the Fourier Transform
of the expression (\ref{a4}).
\bq
&&
\hat T^R[k_\mu k_\nu k_\rho]= \hat T[k_\mu k_\nu k_\rho-
\frac 16(g_{\mu \nu}k_\rho+g_{\mu \rho}k_\nu +g_{\rho \nu}k_\mu)k^2]
\nonumber \\
&& +\frac {1}{12} \left(g_{\mu \nu}(p_\rho +p'_\rho) +
g_{\mu \rho}(p_\nu+ p'_\nu) +g_{\rho \nu}(p_\mu+p'_\mu)  \right)
\nonumber \\
&& \times \left( -\hat B^R[1]((p-p')^2)+ \frac {1}{16 \pi^2}d \right).
\label{a4m}
\eq
As discussed before, the ambiguity is concerned to the traceless part,
and it is due to the
presence of surface terms. We remember that in eq. (\ref{a4}) the trace part is already renormalized and,
in this process, surface terms were discarded. Nevertheless, in the present form, there is no ambiguity in
the trace part and the adjustment of the constant $d$ is done considering this fact.
So, $d$ must be adjusted so as to cancel the
surface terms coming from the traceless part. We can write
\bq
&&\hat T^R[k_\mu k_\nu k_\rho]= J_{\mu \nu \rho}
- \frac 16 \left( g_{\mu \nu}I_\rho(p,p')+g_{\mu
\rho}I_\nu(p,p')+g_{\nu \rho}I_\mu(p,p') \right)   \nonumber \\
&& +\frac {1}{12} \left(g_{\mu \nu}(p_\rho +p'_\rho)+g_{\mu
\rho}(p_\nu+ p'_\nu) +g_{\rho \nu}(p_\mu+p'_\mu)  \right)
\frac {1}{16 \pi^2}d  + \mbox{non ambiguous terms}
\label{a4m2}
\eq
Let us consider the first integral,
\bq
&&J_{\mu \nu \rho}= \int_k^\Lambda \frac{k_\mu k_\nu
k_\rho}{(k^2-m^2)[(p'-k)^2-m^2][(p-k)^2-m^2]}
=2(p+p')^\sigma \int_k^\Lambda \frac{k_\mu k_\nu k_\rho
k_\sigma}{(k^2-m^2)^4} \nonumber \\
&&+ \int_k \frac{k_\mu k_\nu k_\rho (p'^2-2p' \cdot
k)^2}{(k^2-m^2)^3[(p-k)^2-m^2][(p'-k)^2-m^2]}
+ \int_k \frac{k_\mu k_\nu k_\rho (p^2-2p \cdot
k)^2}{(k^2-m^2)^4[(p-k)^2-m^2]} \nonumber \\
&& +\int_k \frac{k_\mu k_\nu k_\rho (p'^2-2p' \cdot k)(p^2-2p \cdot
k)^2}{(k^2-m^2)^4[(p-k)^2-m^2]}\nonumber \\
&&=2(p+p')^\sigma \int_k^\Lambda \frac{k_\mu k_\nu k_\rho
k_\sigma}{(k^2-m^2)^4}
+\mbox{non ambiguous terms},
\eq
which was expanded with the use of the identity (\ref{ident}) and where
we discarded the terms with odd
integrand in $k$. The infrared cutoff $m^2$ disappear in the end. The
divergent integral can be written
in function of surface terms. Let us define
\be
\alpha_2 g_{\mu \nu} \equiv \int^{\Lambda}_k
\frac{g_{\mu\nu}}{(k^2-m^2)^2}-
4\int^{\Lambda}_k
\frac{k_{\mu}k_{\nu}}{(k^2-m^2)^3}
\label{CR1}
\ee
and
\bq
\alpha_3 g_{\{\mu \nu}g_{\alpha \beta\}}  & \equiv &
g_{\{\mu \nu}g_{\alpha \beta \}}
\int^{\Lambda}_k
\frac{1}{(k^2-m^2)^2}
-24\int^{\Lambda}_k
\frac{k_{\mu}k_{\nu}k_{\alpha}k_{\beta}}{(k^2-m^2)^4}.
\label{CR2}
\eq
The parameters ,$\alpha_2$ and $\alpha_3$, are surface terms. It can be
easily shown that
\be
\alpha_2 g_{\mu \nu}= \int_k ^\Lambda \frac{\partial}{\partial k^\mu}
\left( \frac{k_ \nu}{(k^2-m^2)^2} \right)
\ee
and
\be
\int_k^\Lambda \frac{\partial}{\partial k^\beta}
\left[ \frac{4k_\mu k_\nu k_\alpha}{(k^2-m^2)^3} \right]
=g_{\{\mu \nu}g_{\alpha \beta\}}(\alpha_3-\alpha_2).
\ee
These surface terms can be calculated by means of symmetric
integration, with the substitutions
$k_\mu k_\nu \to g_{\mu \nu} k^2/4$ and $k_\mu k_\nu k_\alpha k_\beta
\to g_{\{\mu \nu}g_{\alpha \beta\}}k^4/24$.
We obtain
\be
\alpha_2=-\frac {i}{32 \pi^2} \,\,\,\,\,\, \mbox{and} \,\,\,\,\,\,
\alpha_3= \frac{5i}{96 \pi^2}.
\ee
Returning to the integral, we have
\bq
&&J_{\mu \nu \rho}= -\frac {1}{12}(p+p')^\sigma g_{\{\mu \nu}g_{\rho
\sigma\}} \alpha_3 +
\cdots \nonumber \\
&&= -\frac {1}{12}(p+p')^\sigma g_{\{\mu \nu}g_{\rho \sigma\}}
\frac{5i}{96 \pi^2}
+ \mbox{non ambiguous terms}.
\label{surf1}
\eq
For The integral $I_\mu(p,p')$, it can be written
\bq
&&I_\mu=-2(p+p')^\nu \int_k^\Lambda \frac{k_\mu k_\nu}{(k^2-m^2)^3}+
\cdots  \nonumber \\
&& =\frac 12 (p+p')^\nu g_{\mu \nu} \alpha_2 + \cdots  = -\frac {i}{64
\pi^2}(p+p')^\nu g_{\mu \nu}
+\mbox{non ambiguous terms}.
\label{surf2}
\eq
When the results of the equations (\ref{surf1}) and (\ref{surf2}) are
inserted into the expression
(\ref{a4m2}), and $d$ is chosen to cancel these surface terms, it is
found that $d=\frac i3$, as
expected.

The momentum-space version of eq. (\ref{a1}), taking into account that
$f=-g$ and $\mu=0$, is written bellow:
\be
\hat B[k_\mu k_\nu]= \frac 13 \left( p_\mu p_\nu- \frac 14 g_{\mu \nu}
p^2 \right) I
+ \frac{1}{16 \pi^2}f (p_\mu p_\nu -g_{\mu \nu}p^2),
\label{a1m}
\ee
It is important to call the reader's attention to the fact that the
decomposition in a traceless plus a
trace part was not applied to this basic function. Instead, a
differential equation was solved for $x \neq 0$
in order to find the term on $B[1]$. When the decomposition is
performed, the traceless part is responsible
for an ambiguity due to surface terms (ST), and a local term is added
in order to take care of this problem.
For the present case, there is also an ambiguity, but it is not due to
the first term (on $B[1]$), which is
already free from the ST. It emerges from
the solution of the differential equation. So, the constant $f$ is not
the counterterm to cancel
the surface term. It is actually the local term that survives after the
surface terms were eliminated.
CDR assures it with the use of some consistency equations obtained with
the help of Leibniz rule.
In the case of IR, the same result is achieved by explicit calculation,
as long as ST are discarded.
Let us expand it with the help of the identity (\ref{ident}) ($m^2$ will be set to zero in the final result):
\bq
&& \hat B[k_\mu k_\nu]= I_{\mu \nu}=  \int_k^\Lambda
\frac{k_\mu k_\nu}{(k^2-m^2)[(p-k)^2-m^2]} \nonumber \\
&&= \int_k^\Lambda \frac{k_\mu k_\nu}{(k^2-m^2)}
\left\{ \frac {1}{(k^2-m^2)}-
\frac{p^2-2p \cdot k}{(k^2-m^2)^2}+\frac{(p^2-2p \cdot
k)^2}{(k^2-m^2)^3} \right. \nonumber \\
&& \left. -\frac{(p^2-2p \cdot k)^3}{(k^2-m^2)^3[(p-k)^2-m^2]} \right
\} .
\eq
By discarding the terms with odd integrand and remembering that we can
use a parametrization so that
the quadratic divergence is proportional to $m^2$, we obtain
\bq
I_{\mu \nu}= -p^2\int_k^\Lambda \frac{k_\mu k_\nu}{(k^2-m^2)^3}+
4p^\alpha p^\beta \int_k^\Lambda \frac{k_\mu k_\nu k_\alpha
k_\beta}{(k^2-m^2)^4} +
\mbox{finite},
\eq
in which we can use equations (\ref{CR1}) and (\ref{CR2}) to write
\bq
I_{\mu \nu}= \frac 13 \left( p_\mu p_\nu- \frac 14 g_{\mu \nu} p^2
\right)I_{log}(m^2) +
\mbox{ST}\,\,+ \mbox{finite terms}.
\eq
The scale relation (\ref{scale}), the elimination of surface terms and
the calculus of the finite part
yields
\bq
&&I_{\mu \nu}= \frac 13 \left( p_\mu p_\nu- \frac 14 g_{\mu \nu} p^2
\right)
\left\{ I_{log}(\lambda^2)-b
\ln{\left(-\frac{p^2}{e^2\lambda^2}\right)}\right\}    \nonumber \\
&&- \frac{1}{18}b (p_\mu p_\nu -g_{\mu \nu}p^2),
\eq
where $b=i/(4 \pi)^2$. Then the constant $f$ is found to be the same as
the one fixed by CDR.

In the analysis carried out in this section, we have verified that all
the rules and characteristics of
Constrained Differential Regularization can be mapped in the steps of
Implicit Regularization. The only reason
we have preferred to restrict ourselves, in the examples which we worked out, to the massless case is its
simplicity, since all the features are present.
The same procedure can be applied to the massive case, even when the
problem involves particles with
different masses. In this case, the expressions are greater and the
whole process is more tedious, but it does not
emerge any new feature. Besides, these two methods were applied to a
large sort of problems, massive and
non-massive, always yielding equivalent results.

\section{Concluding Comments}

\indent
The equivalence between Constrained Implicit Regularization (CIR) and
Constrained Differential Renormalization (CDR)
has been analyzed in this paper. The two methods have been tested, with
positive and equivalent results,
in many nontrivial situations, from the symmetry point of view.
The physical dimension of the theory to be treated is not modified
and, for this reason, these methods are candidates to be good tools in
supersymmetric calculations.
In the analysis carried out in this work, it has been shown that each
one of the rules of CDR,
in position-space, have its counterpart in momentum-space, materialized
in one of the rules of CIR.
The relation has been shown to be one to one.
The main characteristic of the two frameworks is the extension of
properties of regular
mathematical objects to the regularized ones. This is accomplished with
the help
of symmetry restoring counterterms. In practice it is very simple, as
long as it is
implemented by a set of rules.

The principles of CIR are being successfully applied at higher order
calculations \cite{ir16}. Differential Renormalization at higher order has been used in
various situations. We believe that the equivalence between these two frameworks
at all orders could also be shown.


\begin{thebibliography}{99}



\bibitem{dimr} G. 't Hooft and M. Veltman, \textit{{Nucl. Phys.
\textbf{B}}}, \textbf{44},
189 (1972).

\bibitem{dimred} W. Siegel, \textit{{Phys. Lett. \textbf{B}}}
\textbf{84}, 193 (1979);
D. M. Capper, D. R. T. Jones and P. van Nieuwenhuizen, \textit{{Nucl.
Phys. \textbf{B}}}
\textbf{167}, 479 (1980)

\bibitem{stock} Dominik St\"ockinger, hep-ph/0602005

\bibitem{df1} D. Z. Freedman, K. Johnson and J. I. Latorre,
\textit{{Nucl. Phys. \textbf{B}}}
\textbf{371}, 353 (1992)

\bibitem{df2} P. E. Haagensen, \textit {{ Mod. Phys. Lett.\textbf{A}}}
\textbf{7}, 893 (1992);
R. Mu\~noz-Tapia, \textit {{ Phys. Lett.\textbf{B}}} \textbf{295}, 95
(1992);
D. Z Freedman, G. Grignani, K. Johnson and N. Rius, \textit {{ Ann.
Phys. }} \textbf{218}, 75 (1992);
P. E. Haagensen, J. I. Latorre, \textit {{ Ann. Phys.}} \textbf{221},
77 (1993);
C. Manuel, \textit {{Int. J. Mod. Phys.\textbf{A}}} \textbf{8}, 3223
(1993);
D. Z. Freedman, G. Lozano and N. Rius, \textit {{Phys. Rev.
\textbf{D}}} \textbf{49}, 1054 (1994);
J. Comellas, P.E. Haagensen and J. I. Latorre, \textit {{Int. J. Mod.
Phys.\textbf{A}}} \textbf{10}, 2819 (1995);
M. Chaichian, W. F. Chen, H. C. Lee, \textit {{Phys.Lett. \textbf{B}}}
\textbf{409}, 325 (1997);
V. A. Smirnov, \textit {{Int. J. Mod. Phys.\textbf{A}}} \textbf{12},
4241 (1997);
D. Anselmi, D. Z. Freedman, M. T. Grisaru, A. A. Johansen,
hep-th/9708042.

\bibitem{df3} J. I. Latorre, C. Manuel and X. Vilasis-Cardona, \textit
{{Ann. Phys.}} \textbf{231}, 141 (1994);
G. Dunne, N. Rius, \textit {{Phys. Lett. \textbf{B}}} \textbf{293}, 367
(1992);
V. A. Smirnov, \textit {{Nucl. Phys.\textbf{B}}} \textbf{427}, 325
(1994).

\bibitem{df4} D. Z. Freedman, K. Johnson, R. Mu\~noz-Tapia and X.
Vilasis-Cardona,
\textit {{Nucl. Phys.\textbf{B}}} \textbf{395}, 454 (1993).

\bibitem{df5} P. E. Haagensen, J. I. Latorre, \textit {{Phys. Lett.
\textbf{B}}} \textbf{283}, 293 (1992).

\bibitem{cdf1} F. del Aguila, A. Culatti, R. Mu\~noz Tapia and M.
P\'erez Victoria,
\textit{{Phys. Lett. \textbf{B}}} \textbf{419}, 263 (1998).

\bibitem{cdf2} M. P\'erez Victoria, \textit{{Phys. Lett. \textbf{B}}}
\textbf{442}, 315 (1998)

\bibitem{cdf3} F. del Aguila, A. Culatti, R. Mu\~noz Tapia and M.
P\'erez Victoria,
\textit{{Nucl. Phys. \textbf{B}}}
\textbf{504}, 532 (1997)

\bibitem{cdf4} F. del Aguila, A. Culatti, R. Mu\~noz Tapia and M.
P\'erez Victoria,
hep-ph/9711474

\bibitem{cdf5} F. del Aguila, A. Culatti, R. Mu\~noz Tapia and M.
P\'erez Victoria,
\textit{{Nucl. Phys. \textbf{B}}}
\textbf{537}, 561 (1999)

\bibitem{cdf6} F. del Aguila, M. P\'erez Victoria, hep-ph/9901291

\bibitem{cdf7} F. del Aguila, M. P\'erez-Victoria, \textit{{Acta
Phys.Polon.\textbf{B}}} \textbf{28},2279 (1997)

\bibitem{ir1} O. A. Battistel, {\it PhD thesis}, Federal University of
Minas Gerais (2000)

\bibitem{ir2} O. A. Battistel, A. L. Mota, M. C. Nemes \textit{{Mod.
Phys. Lett. \textbf{A}}}
\textbf{13} 1597 (1998)

\bibitem{ir3}  A. P. Ba\^{e}ta Scarpelli, O. A. Battistel and M. C.
Nemes,
\textit{{Braz. J. Phys.}}\textbf{28}, 161 (1998).
\textbf{13} 1597 (1998)

\bibitem{ir4}  A. P. Ba\^{e}ta Scarpelli, M. Sampaio and M. C. Nemes,
\textit{%
{Phys. Rev. \textbf{D}}} \textbf{63}, 046004 (2001)

\bibitem{ir5}  A. P. Ba\^{e}ta Scarpelli, M. Sampaio, B. Hiller and M.
C. Nemes, \textit{{Phys. Rev. \textbf{D}}} \textbf{64}, 046013 (2001)

\bibitem {ir6} M. Sampaio, A. P. Ba\^eta Scarpelli, B. Hiller, A.
Brizola,
M. C. Nemes and S. Gobira, \textit{%
{Phys. Rev. \textbf{D}}} \textbf{65}, 125023 (2002)

\bibitem{ir7} S. R. Gobira and M. C. Nemes, \textit{ Int. J. Theor.
Phys.} \textbf{42},
2765 (2003)

\bibitem{ir8} D. Carneiro, A. P. Ba\^eta Scarpelli, M. Sampaio and M.
C.
Nemes,
\textit{JHEP} \textbf{12}, 044 (2003)

\bibitem{ir9} M. Sampaio, A. P. Ba\^eta Scarpelli, J. E. Ottoni, M. C.
Nemes, \textit{{Int.J.Theor.Phys}} \textbf{45}, 436 (2006)

\bibitem{ir10} Leonardo A.M. Souza, Marcos Sampaio, M.C. Nemes,
\textit{{Phys. Lett. \textbf{B}}} \textbf{632}, 717 (2006)

\bibitem{ir11} Carlos. R. Pontes, A. P. Ba\^eta Scarpelli, Marcos
Sampaio, M. C.
Nemes,
{\it Implicit regularization beyond one loop order: scalar field
theories},
hep-th/0605116

\bibitem{ir12}J. E. Ottoni, A. P. Ba\^eta Scarpelli, Marcos Sampaio, M.
C. Nemes,
\textit{{Phys. Lett. \textbf{B}}} \textbf{642}, 253 (2006).

\bibitem{ir13} E. W. Dias, B. Hiller, A. L. Mota, M. C. Nemes and
Marcos Sampaio,
\textit{{Mod. Phys. Lett. \textbf{A}}} \textbf{21}, 339 (2006).

\bibitem{ir14} B. Hiller, A. L. Mota, M. C. Nemes, A. A. Osipov and
Marcos Sampaio,
\textit{{Nucl. Phys. \textbf{A}}} \textbf{769}, 53 (2006).

\bibitem{ir15} O. A. Battistel, G. Dallabona,
\textit{{Eur. Phys. J. \textbf{C}}} \textbf{45}, 721 (2006).

\bibitem{ir16} E. W. Dias, A. P. Ba\^eta Scarpelli, Marcos Sampaio, M.
C. Nemes,
{\it Implicit regularization beyond one loop order: gauge field
theories}, work in progress.

\bibitem{ABJ} S. Adler, \textit{{Phys. Rev.}} \textbf{177}, 2426
(1969);
J. S. Bell and R. Jackiw, \textit{{Nuovo Cimento}} \textbf{51}, 47
(1969)

\bibitem{cdf8} C. Seijas, hep-th/0604071

\bibitem{Mackeon} V. Elias, G. McKeon, S. B. Phillips and R. B. Mann,
\textit{{Phys. Lett. \textbf{B}}} \textbf{133}, 83 (1983)

\bibitem{Bogoliubov} N.N. Bogoliubov, A. A. Logunov, I. T. Todorov,
{\it Introduction to Axiomatic Quantum Field Theory}, (Benjamim Cummings), Massachussets, (1975)

\end{thebibliography}
\end{document}